\documentclass[a4paper,12pt]{article}

\newcommand{\be}{\begin{equation}}
\newcommand{\ee}{\end{equation}}

\newcommand{\f}{\frac}

\newcommand{\calO}{{\mathcal O}}

\newcommand{\spose}[1]{\hbox to 0pt{#1\hss}}
\newcommand{\lta}{\mathrel{\spose{\lower 3pt\hbox{$\mathchar"218$}}
 \raise 2.0pt\hbox{$\mathchar"13C$}}}
\newcommand{\gta}{\mathrel{\spose{\lower 3pt\hbox{$\mathchar"218$}}
 \raise 2.0pt\hbox{$\mathchar"13E$}}}

\begin{document}

\author{Micha{\l} J.~Chodorowski}

\title{Is space really expanding? A counterexample}
\date{\small \it Copernicus Astronomical Center, Bartycka 18, 00--716 
Warsaw, Poland }
\maketitle

\begin{abstract}
\normalsize 
In all Friedman models, the cosmological redshift is widely
interpreted as a consequence of the general-relativistic phenomenon of
{\em expansion of space}. Other commonly believed consequences of this
phenomenon are superluminal recession velocities of distant galaxies,
and the distance to the particle horizon greater than $c\,t$ (where
$t$ is the age of the Universe), in apparent conflict with special
relativity. Here, we study a particular Friedman model: empty
universe. This model exhibits both cosmological redshift, superluminal
velocities and infinite distance to the horizon. However, we show that
the cosmological redshift is there simply a relativistic Doppler
shift. Moreover, apparently superluminal velocities and `acausal'
distance to the horizon are in fact a direct consequence of
special-relativistic phenomenon of time dilation, as well as of the
adopted definition of distance in cosmology. There is no conflict with
special relativity, whatsoever. In particular, {\em inertial\/}
recession velocities are subluminal. Since in the real Universe,
sufficiently distant galaxies recede with relativistic velocities,
these special-relativistic effects must be at least partly responsible
for the cosmological redshift and the aforementioned
`superluminalities', commonly attributed to the expansion of
space. Let us finish with a question resembling a Buddhism-Zen `koan':
in an empty universe, what is expanding?
\end{abstract}

\section{Introduction}
\label{sec:intro}
What is the physical interpretation of the cosmological redshift? It is
well known that, although in general ``the redshift cannot be thought
of as a {\em global\/} Doppler shift, it is correct to think of the
effect as an accumulation of the infinitesimal Doppler shifts caused
by photons passing between fundamental observers separated by a small
distance'' (Peacock 1999). In all Friedman cosmological models, the
Universe is isotropic and homogeneous for a privileged set of
so-called Fundamental Observers (FOs) co-moving with matter, all
measuring the same cosmic time. The cosmological redshift can be thus
thought of as a result of relative motions of the FOs along the
photon's trajectory.

An alternative, and widespread, interpretation of the cosmological
redshift is that it is a direct consequence of the
general-relativistic phenomenon of {\em expansion of space}. In all
homogeneous and isotropic (i.e., Friedman) cosmological models,
distances between various galaxies grow by the same factor, called the
scale factor, $a$, which is a function of only the cosmic time,
$\tau$: $a = a(\tau)$. The wavelength of the emitted photon,
$\lambda$, undergoes redshift, $z$, which is simply related to the
values of the scale factor at the time of emission, $\tau_e$, and
observation, $\tau_o$,

\be
\frac{\lambda_o}{\lambda_e} \equiv 1 + z = \frac{a(\tau_o)}{a(\tau_e)} .
\label{eq:red}
\ee 
A seemingly natural interpretation of the above equation is that space
expands, and the wavelengths of photons grow accordingly in time. But
why is then, say, Brooklyn, {\em not\/} expanding? Quoting Lineweaver
\& Davis (2005), ``In `Annie Hall', the movie character played by the
young Woody Allen explains to his doctor and mother why he can't do
his homework. `The universe is expanding\ldots The universe is
everything, and if it's expanding, someday it will break apart and
that would be the end of anything!'. But his mother knows better:
`You're here in Brooklyn. Brooklyn is not expanding!'{''}
Certainly. But how can a tiny photon partake in the global expansion
of the universe, if something as big as Brooklyn does not?

In his textbook `Cosmological Physics', John Peacock calls the idea of
expanding space ``perhaps the worst misconception about the big
bang''. ``Many semi-popular accounts of cosmology contain statements
to the effect that `space itself is swelling up' in causing the
galaxies to separate. This seems to imply that all objects are being
stretched by some mysterious force: are we to infer that humans who
survived for a Hubble time would find themselves to be roughly four
meters tall? Certainly not. Apart from anything else, this would be a
profoundly anti-relativistic notion, since relativity teaches us that
properties of objects in local inertial frames are independent of the
global properties of spacetime.'' However, expanding space is a deeply
rooted myth (e.g., Mc Vittie 1934; Einstein \& Straus 1945; Einstein
\& Straus 1946) and such myths die hard. The following properties of
the Friedman models are commonly attributed to general-relativistic
expansion of space:
\begin{enumerate}
\item Cosmological redshift;
\item Superluminal recession velocities of distant galaxies;
\item Distance to the horizon greater than $c\,t$, where $c$ is the
speed of light and $t$ is the age of the Universe.
\end{enumerate}
Expansion of space is regarded as general-relativistic because the
properties~2.\ and~3.\ seem to be in sharp conflict with special
relativity. In particular, with respect to property 3., it is argued
that photons travel locally with the speed of light, but space
expands, providing an additional stretching of the distance. With
respect to property 2., we can often read that ``the velocity in
Hubble's law is a recession velocity caused by the expansion of space,
not a motion through space. It is a general-relativistic effect and is
not bound by the special-relativistic limit'' (Lineweaver \& Davis
2005).

In this paper, as a counterexample to the idea of expanding space we
study the model of an empty universe. It is a rather particular
Friedman model, since it is devoid of matter, in a classical
sense. However, we will see that it exhibits all properties 1.--3. It
is a good counterexample to the idea of expanding space because its
spacetime is simply the Minkowskian spacetime of special relativity
(SR). The spacelike section of the Minkowskian spacetime is Euclidean
(i.e., flat) {\em static\/} space. In the empty model, expanding are
only fictitious, massless (so non-interacting) FOs, all in constant
relative motion. Without them, there would be no expansion at
all. After all, how to define motion without any object of reference?
This suggests that what really matters is the cosmic substratum (here,
massless by assumption) and its relative motions.

One can make a coordinate transformation, transforming the underlying
metric of the empty model from its Minkowskian form to the form
expressed in the local coordinates of FOs (Peacock 1999; Longair
2003). Then it turns out to be an open Friedman model: its spacelike
section is negatively curved space of hyperbolic geometry, evolving in
time. There is no absolute space already in SR. However, according to
any inertial observer, space is flat and static. We see that in
general relativity (GR), even the curvature of space\footnote{Unlike
the curvature of whole spacetime.} or its dynamical state (static or
evolving) is not an invariant of an arbitrary coordinate
transformation. On the other hand, using the latter form of the metric
it is straightforward to derive properties 1.--3.\ of the empty
model. On this basis one may still argue that although
general-relativistic expansion of space is not absolute, it is a fact
in the privileged coordinates of FOs.

To show that even this line of reasoning is incorrect, here we derive
properties 1.--3.\ of the empty model using only special-relativistic
concepts. In particular, we do not use the notion of a metric at
all. We demonstrate that in this model, the cosmological redshift is a
relativistic Doppler shift. Furthermore, we show that the conflict of
properties~2.\ and~3\ with SR is only apparent. In fact, they are a
direct consequence of special-relativistic phenomenon of time
dilation, as well as of the adopted definition of distance in
cosmology. In other words, at least in the empty model, properties
1.--3.\ are in accordance with SR, and are fully explicable as the
results of real motions in space. Therefore, at least in this case,
the existence of these properties cannot be regarded as an argument
for general-relativistic expansion of space. Alternatively, there is
at least one Friedman model, in which expansion of space, in
detachment from expanding matter, is certainly an illusion.

This paper is organized as follows. In Section~\ref{sec:redshift} we
show that in the empty model, the origin of the cosmological redshift
is entirely kinematic. In Section~\ref{sec:cmb} we derive the
temperature of the cosmic microwave background at redshift $z$. In
Section~\ref{sec:super} we derive a formula for the recession
velocities of distant objects. In Section~\ref{sec:horizon} we study
the distance to the particle horizon. Conclusions are in
Section~\ref{sec:conc}.

\section{The cosmological redshift}
\label{sec:redshift}
What happens to a photon traveling through an empty universe? A simple
(and correct) answer is that nothing.  Quoting the lyrics of a song by
Grzegorz Turnau, ``in fact, nothing happens and nothing occurs till
the very end''.\footnote{In the original: ``tak naprawd\c e nie dzieje
si\c e nic i nie zdarza si\c e nic, a\.z do ko\'nca''.} The end occurs
when the emitted photon is finally absorbed by an observer's eye, a
photographic plate, or a CCD device. This absorption reveals that the
frequency of the photon is redshifted. The only possible
interpretation of this redshift is a Doppler shift, due to relative
motion of the emitter and the absorber.

In an empty universe, these motions can be described entirely by means
of the Milne kinematic model. In this model, the cosmic arena of
physical events is the pre-existing Minkowski spacetime. In the origin
of the coordinate system, $O$, at time $t = 0$ an `explosion' takes
place, sending radially Fundamental Observers (FOs) with constant
velocities in the range of speeds $(0,c)$. Let's place a source of
radiation at the origin of the coordinate system.\footnote{An
analogous calculation with the observer at the origin yields the same
result.} At time $t_e$ the source emits photons, which at time $t_o$
reach a FO moving with velocity $v$, such that \be v t_o = c (t_o -
t_e) .
\label{eq:v}
\ee 
The observer sees them redshifted, due to the Doppler effect. The
special-relativistic formula for the Doppler effect is
\be
1 + z = \left(\f{1 + \beta}{1 - \beta}\right)^{1/2} ,
\label{eq:redshift}
\ee
where $z$ is the photons' redshift and $\beta \equiv v/c$.
From Eq.~(\ref{eq:v}) we have 
\be \beta = 1 - {p}^{-1} ,
\label{eq:beta_M}
\ee 
where ${p} \equiv t_o/t_e$, hence
\be
1 + z = (2 {p} - 1)^{1/2} .
\label{eq:1+z}
\ee 
We emphasize that time $t$ is measured in the inertial frame of the
source, i.e., by a set of synchronized clocks (with that at the
origin), remaining in rest relative to it. The observer traveling
with velocity $v$ relative to the source, carries his own clock which
shows his {\em proper\/} time, $\tau$. This clock {\em delays\/}
relative to time $t$:
\be
t = \gamma(v) \tau \,,
\label{eq:dilat}
\ee
where $\gamma(v) = (1 - \beta^2)^{-1/2}$; hence $t_o = \gamma(v)
\tau_o$. On the other hand, the clock at the source measures its own
proper time, hence $t_e = \tau_e$. This yields
\be 
{p} = \frac{t_o}{t_e} = \gamma(v) \frac{\tau_o}{\tau_e} ,
\label{eq:t_ratio}
\ee 
The instants of time $\tau_e$ and $\tau_o$ are measured by the FOs
respectively at the point of emission and observation of photons, so
they are the instants of {\em cosmic\/} time.

From Equation~(\ref{eq:beta_M}) we have $\gamma(v) = \left[p^{-1} (2 -
p^{-1}) \right]^{-1/2}$, or
\be
p = \left[p^{-1} (2 - p^{-1}) \right]^{-1/2} \frac{\tau_o}{\tau_e} .
\label{eq:p_ratio}
\ee 
Solving this equation for $p$ we obtain
\be 
p = \frac{1}{2} \left[1 + \left(\frac{\tau_o}{\tau_e}\right)^2 \right] .
\label{eq:p_fin}
\ee 
Using the above in Equation~(\ref{eq:1+z}) yields finally
\be
1 + z = \frac{\tau_o}{\tau_e} .
\label{eq:z}
\ee 
The scale factor, $a$, of the Robertson-Walker form of the metric for
an empty universe grows linearly with time: $a(\tau) =
\tau/\tau_o$. For this form of the metric, the cosmological redshift
is in general $1 + z = a(\tau_o)/a(\tau_e)$, so for an empty universe,
$1 + z = \tau_o/\tau_e$. Equation~(\ref{eq:z}), derived using only
special-relativistic concepts, coincides with this formula. We see
thus that in the empty model, the origin of the cosmological
redshift is entirely kinematic (Peacock 1999; Chodorowski 2005; see
also Whiting 2004). 

\section{Temperature of the CMB}
\label{sec:cmb}
One can sometimes hear that the temperature of the cosmic microwave
background (CMB), measured at redshift $z$, is a strong observational
evidence for expansion of space. For all Friedman models, this
temperature is predicted to be $(1 + z) T_o$, where $T_o$ is its
present value, as indeed observed (Srianand, Petitjean \& Ledoux
2000). If only matter expanded, this temperature would be expected to
be just $T_o$ (Bajtlik, private communication). To see that the last
statement is wrong, let us analyze the redshift of photons in the rest
frame of the source. At time $t_d$, corresponding to the time of
decoupling of photons from matter, the source emits photons of
temperature $T_d$ towards the observer. At time $t_c$ they reach a
molecular cloud. Some of them are absorbed by the cloud, providing a
thermal bath for its atoms and molecules, of temperature $T_c$. The
remaining photons are not disturbed and reach the observer at time
$t_o$. What is the value of the temperature $T_c$? In the empty model,
the redshift of unabsorbed photons, $z_{\rm CMB}$, is related to the
velocity of the observer relative to the source, $\beta_o$, by
Equation~(\ref{eq:redshift}).  Similarly, Equation~(\ref{eq:redshift})
also relates the redshift of the cloud relative to the source, $z_c$,
to its velocity, $\beta_c$. The redshift of the observer {\em relative
to the cloud,} or the cloud relative to the observer, is
\be 
1 + z = \left(\f{1 + \beta'}{1 - \beta'}\right)^{1/2} ,
\label{eq:red'}
\ee
where $\beta'$ is the relative velocity of the observer and the
cloud. From the special-relativistic law of addition of velocities, 
\be
\beta' = \frac{\beta_o - \beta_c}{1 - \beta_o \beta_c} .
\label{eq:beta'}
\ee
We thus have
\be 
(1 + z)^2 = \frac{1 - \beta_o\beta_c + \beta_o -\beta_c}{1 -
\beta_o\beta_c - \beta_o + \beta_c} = \frac{(1 + \beta_o) (1 -
\beta_c)}{(1 - \beta_o) (1 + \beta_c)} = \frac{(1 + z_{\rm
CMB})^2}{(1 + z_c)^2} ,
\label{eq:red^2}
\ee
or 
\be 
1 + z_c = \frac{1 + z_{\rm CMB}}{1 + z} .
\label{eq:z_c}
\ee
Since $T_c = T_d/(1 + z_c)$, the above equation yields
\be
T_c = \frac{1 + z}{1 + z_{\rm CMB}} T_d = (1 + z) T_o .
\label{eq:T_c}
\ee
In other words, in the empty model the {\em local\/} temperature of
the CMB photons at a source of redshift $z$ is $(1 + z) T_o$, in
agreement with GR. Specifically, Equation~(\ref{eq:T_c}) immediately
follows from Equation~(\ref{eq:red}):
\be
1 + z_{\rm CMB} = \frac{a_o}{a_c} \frac{a_c}{a_e} = (1 + z) (1 + z_c), 
\label{eq:z_CMB}
\ee 
coinciding with Equation~(\ref{eq:z_c}). Thus the local coordinates
of FOs are more convenient for calculations than the Minkowskian
coordinates, which we have used. However, we have obtained the same
result. This is not surprising since Equation~(\ref{eq:T_c}) involves
three observables: $T_c$, $T_o$, and $z$. Regardless of specific
definitions of coordinates in a given coordinate system, their
consistent application should lead to the same result in terms of
observables. 

As a corollary of this section, the temperature of the CMB photons at
redshift $z$ is a strong observational evidence for expanding
universe, but, at least in the empty model, it is fully consistent
with real motions of matter.

\section{Superluminal recession velocities}
\label{sec:super}
In an empty universe one can use {\em global\/} Minkowskian
coordinates of distance and time, $r$ and $t$. We recall that FOs
emanate from the origin $r = 0$ at $t = 0$, and travel with constant
velocities. The trajectory of a FO is simple in Minkowskian
coordinates: $r = v_M t$, where $v_M$ is its Minkowskian velocity. 
(Analysis in the preceding section involved Minkowskian
velocities.) {\em Any\/} velocity of any FO is calculated {\em along
its trajectory.\/} The Minkowskian velocity of a given FO is
\be 
v = \frac{d r}{d t}\biggr |_{r = v_M t}\! = \frac{d (v_M t)}{d t} = v_M .
\label{eq:v_Mink}
\ee
Minkowskian coordinates define what Milne called `private space'
of an observer. 

However, in cosmology we usually measure distances on the hypersurface
of constant {\em proper\/} time, $\tau$, of all FOs. Along the line of
sight to a distant object we consider a hypothetical series of closely
spaced FOs, and as the distance we adopt a sum of all distances
measured by them to their nearest neighbours. In Milne's terminology,
this measurement is performed in `public space'. Since all FOs are
in motion relative to the observer at $r=0$, the distances and time
they measure are subject to special-relativistic phenomena of
length-contraction and time-dilation. For a given FO at the point $r$
in the moment $t$, his clock shows time $\tau$, that is dilated
relative to time $t$. Using Equation~(\ref{eq:dilat}), $t =
\gamma(v_M)\tau$, or
\be
t = \gamma(r/t) \tau .
\label{eq:dilat_2}
\ee
We recall that time $t$ is measured in the inertial frame of the FO
{\em at the origin,\/} i.e., that is measured by a set of synchronized
clocks (with that at the origin), remaining in rest relative to
it. Similarly, the observer at $r$ measures the distance $d l$ (at
$\tau$) that is different from $d r$ measured by the central
FO. Specifically, \be dl_{|\tau} = \frac{dr_{|\tau}}{\gamma(r/t)} .
\label{eq:contract}
\ee 
Note that a hypothetical ruler of length $dl$ is stationary not in the
frame $O(r,t)$, but in the frame $O'(r',t'=\tau)$, so the above
equation may appear at first sight in contradiction with classical
special-relativistic length contraction of a moving body. The reason
why in this equation, derived from the Lorentz transformation between
the two frames, the distance divided by $\gamma$ is not $dl$ but $dr$
is that these distances are measured not at constant $t$, but at
constant $\tau$ (Peacock 1999). In fact, Equation~(\ref{eq:contract})
{\em is\/} the classical length-contraction formula for the frames $O$
and $O'$ with exchanged roles.

From Equation~(\ref{eq:dilat}) we obtain $t^2 = \tau^2 +
r^2/c^2$. Substituting this in Equation~(\ref{eq:contract}) yields
\be
dl_{|\tau} = \frac{dr_{|\tau}}{\sqrt{1 + r^2/c^2 \tau^2}} \,.
\label{eq:contract2}
\ee 
Integrating Equation~(\ref{eq:contract2}) for a constant $\tau$ we
have
\be 
l_{|\tau} = \int_{0}^{r}{\frac{d\tilde{r}_{|\tau}}{\sqrt{1 +
{\tilde{r}}^2/c^2 \tau^2}}} \,,
\label{eq:contract3}
\ee 
hence (Longair 2003)
\be
l = c\tau \sinh^{-1}(r/c\tau) .
\label{eq:l(r)}
\ee 
Thus, `public-space' distance, $l$, is length-contracted compared to
`private-space' distance, $r$. After the proper-time interval
$d\tau$, FOs again measure distances to their nearest neighbours, and
the cumulative distance is $l(\tau + d\tau)$. The `public-space'
velocity of the FO with Minkowskian velocity $v_M$ is thus $v_{\rm
rec} = dl/d\tau$, again calculated along its trajectory. We have
\be 
v_{\rm rec} = \frac{d}{d\tau} \left[c\tau \sinh^{-1}(r/c\tau)\right]
\biggr |_{r = v_M t}\! = \frac{d}{d\tau} \left[c\tau
\sinh^{-1}\left(\frac{v_M t}{c\tau}\right)\right] = \frac{d}{d\tau}
\left[c\tau \sinh^{-1}\left(\beta_M \gamma_M\right)\right] ,
\label{eq:v_rec_Milne}
\ee 
where in the last equality we have used Equation~(\ref{eq:dilat});
here $\beta_M = v_M/c$, and $\gamma_M = \gamma(v_M)$. Since in
an empty universe, $v_M$ of a given FO remains constant, we see that
`public-space' distance to this observer ($l$) grows linearly with
`public-time' ($\tau$). Therefore, simply
\be
v_{\rm rec} = c \cdot \sinh^{-1}\left(\beta_M \gamma_M\right)\!:
\label{eq:v_rec_Milne2}
\ee 
also `public-space' velocity of any FO remains constant. 

Redshift of photons emitted by a given FO is related by
Equation~(\ref{eq:redshift}) to its {\em Minkowskian\/} (i.e., inertial)
recession velocity. This yields $\gamma_M = (1 + z)/(1 + \beta_M)$. 
After simple algebra, we obtain
\be
v_{\rm rec} = c \cdot \sinh^{-1}\left[\frac{z(1 + z/2)}{1 + z}\right],
\label{eq:v_rec_Milne3}
\ee 
or
\be
v_{\rm rec} = c\cdot \ln(1 + z).
\label{eq:v_rec_Milne_fin}
\ee 
The last step can be easily verified by showing that
Equation~(\ref{eq:v_rec_Milne_fin}) implies 
\be
\sinh(v_{\rm rec}/c) = \frac{z(1 + z/2)}{1 + z},
\label{eq:identity}
\ee 
thus it reproduces Equation~(\ref{eq:v_rec_Milne3}). 

Solving Equation~(\ref{eq:redshift}) for $v_M$ we obtain
\be
v_M = c\, \frac{(1+z)^2 - 1}{(1+z)^2 + 1} .
\label{eq:v_Dopp}
\ee 
Comparing Equation~(\ref{eq:v_rec_Milne_fin}) with~(\ref{eq:v_Dopp}) we
see that `public-space' recession velocity of any FO is a different
function of redshift than its `private-space' (or inertial)
velocity. These velocities are equal only to second order in redshift
($v \simeq z[1 - z/2 + \calO(z^2)]$), because time-dilation and
length-contraction factors are unity plus terms which are of second
order in $\beta_M$, so in $z$. In almost all Friedman models, objects
with sufficiently large redshifts recede from the central observer
with superluminal velocities (greater than $c$). For example, in an
Einstein-de Sitter universe ($\Omega_m = 1$ and $\Omega_\Lambda = 0$),
the `public-space' recession velocity as a function of redshift is
\be 
v_{\rm rec} = 2\, c\left[1 - (1+z)^{-1/2}\right] ,
\label{eq:v_rec_EdS}
\ee
hence $v_{\rm rec} > c$ for $z > 3$ (Murdoch 1977). In particular, the
velocity of the so-called particle horizon (corresponding to infinite
redshift) is $2c$. In an empty universe, `public-space' recession
velocities are not only superluminal for sufficiently large redshifts;
they are even unbounded. Does it imply violation of special relativity
in cosmology?  Of course not. Apart from anything else, deriving
Equation~(\ref{eq:v_rec_Milne_fin}) we have used nothing except
special relativity! Constancy of the speed of light, and subluminality
of the motion of massive bodies, applies only to {\em inertial\/}
frames. However, `public-space' distance is a hybrid of distances
measured in different inertial frames, all in relative motion. Since
the resulting $v_{\rm rec}$ is not measured in any single inertial
frame, there is no violation of special relativity (Davis 2004).

Specifically, `public-space' distance is measured at constant proper
time of FOs. Time-dilation formula tells us that according to the
central observer, this measurement is done at the instant of time $t_i
= \gamma(v_i) \tau$, where $v_i$ is the Minkowskian velocity of the
$i$-th FO. Since more distant FOs have greater velocities, it is
obvious that for two different FOs, $t_i \ne t_j$. Therefore,
according to the central observer, different (sub)distances are not
measured {\em simultaneously.\/} Simultaneity is a crucial condition
of special-relativistic measurements of distances to and sizes of
bodies in motion. Waiving this condition may have important
consequences and indeed, it does have! The problem with the real
Universe is that it is filled with matter and expanding, so there are
no global inertial frames. Then, measuring distance (along geodesics)
on the hypersurface of constant proper time of FOs is something most
natural to do. We should, however, bear in mind the `costs' of such a
definition of distance. One of them are apparently superluminal
recession velocities of distant galaxies.

\section{Particle horizon}
\label{sec:horizon}
From the Robertson-Walker (RW) form of the metric for an empty
universe it is straightforward to derive the {\em comoving\/} radial
distance, $x \equiv a(\tau_o) l/a(\tau)$, to a source lying at redshift $z$,
\be
x = c H_o^{-1} \ln(1 + z) \,.
\label{eq:comov}
\ee
We note in passing that since $v_{\rm rec} = H_o x$, we have $v_{\rm
rec} = c \ln(1 + z)$, in agreement with
Equation~(\ref{eq:v_rec_Milne_fin}). By definition, $x = l_o =
l(\tau_o)$. Writing $v_{\rm rec} = dl/d\tau$, from
Equation~(\ref{eq:v_rec_Milne_fin}) we immediately obtain
\be
l = c \tau \ln(1 + z) \,,
\label{eq:l(z)}
\ee 
hence $l_o = c \tau_o \ln(1 + z)$. The central observer observes other
FOs receding with constant velocities, so for a given FO, its
Minkowskian distance is $r_o = v t_o$, or $v = H_o r_o$, where $H_o =
t_o^{-1}$: this is the Hubble law in Minkowskian coordinates. The
central observer measures its proper time, so $t_o = \tau_o$. We have
thus $\tau_o = t_o = H_o^{-1}$, hence $l_o = c H_o^{-1} \ln(1 + z)$,
in agreement with Equation~(\ref{eq:comov}). 

Equation~(\ref{eq:red}) for the cosmological redshift shows that the
limit $z \to \infty$ corresponds to the limit $\tau_e \to 0$, so
infinitely redshifted photons were emitted just at the Big Bang. The
current distance to their source is called the {\em particle
horizon\/} (at a given instant of time, we cannot see further
sources). For example, the present value of the particle horizon in an
Einstein-de Sitter universe is $ 2 c H_o^{-1} = 3 c \tau_o$, where
$\tau_o$ is the present age of the universe. This value seems to imply
that the horizon recedes with superluminal velocity. Indeed, in
Section~\ref{sec:super} we have noted that the present value of the
`public space' horizon's velocity in this model is $2 c$ (see
Eq.~\ref{eq:v_rec_EdS}). We will return to this topic later on.

Returning now to the empty model, from Equation~(\ref{eq:l(z)}), the
present value of the particle horizon in an empty universe is
\be
\lim_{z \to \infty} l_o = c H_o^{-1} \lim_{z \to \infty}
\ln(1 + z) = \infty .
\label{eq:limit}
\ee 
Therefore, the empty model does not have the particle horizon, or has
it at infinity. Why? This is a direct consequence of the
special-relativistic phenomenon of time dilation. The present value of
the `public-space' distance to any object is measured at the proper
time $\tau_o$. Time-dilation formula tells us that according to the
central observer, this measurement is done at the instant of time $t_o
= \gamma(v_M) \tau_o$, where $v_M$ is the Minkowskian velocity of the
receding object. The limit $z \to \infty$ implies $v_M \to c$, hence
$t_o \to \infty$. In other words, in the inertial system of the
central observer, a source with $z = \infty$ travels at the speed of
light, so it is infinitely time-dilated, so it needs infinite time $t$
to acquire any non-zero (finite) value of its proper time. Travelling
with the velocity of light, after infinite time it is infinitely far
away, even in terms of the `public-space' distance $l$. 

In the Einstein-de Sitter model, the universe decelerates, so relative
to the central observer, any initially ultra-relativistic source slows
down. This causes `almost'-luminal sources (with $v_M \to c$, so $z
\to \infty$) to become significantly subluminal. This makes the
time-dilation effect finite, or makes the proper time of the FO
sitting on the source to flow. Since it takes finite time $t_o$ to
acquire the value of the proper time $\tau_o$, the distance travelled
is also finite. It is interesting to note that for the currently
favoured cosmological model, $\Omega_m = 0.3$ and $\Omega_\Lambda =
0.7$, the radius of the particle horizon is approximately $3.4 c
\tau_o$ (Davis \& Lineweaver 2004), not much greater than the value
for the Einstein-de Sitter model ($3 c \tau_o$). This is consistent
with the fact that the current acceleration of the Universe started
fairly recently (in this particular model, at $z \simeq 0.7$).

Another model in which the particle horizon is infinite is the de
Sitter model ($\Omega_m = 0$, $\Omega_\Lambda = 1.0$), where $l_o = c
H_o^{-1} z$. A de Sitter universe constantly accelerates, so it is not
surprising that the divergence of $l_o$ as a function of $z$ is
stronger here than in the empty model, since here an
ultra-relativistic source becomes even more ultra-relativistic.

\section{Conclusions}
\label{sec:conc}
In this paper, as a counterexample to the idea of expanding space, we
have studied the dynamics of the empty model. We have shown that the
cosmological redshift is there a result of the real motion of the
source, i.e., a Doppler shift. We have verified that the local
temperature of the CMB photons at a source of redshift $z$ is a factor
of $(1 + z)$ greater than its present value, in agreement with GR. We
have shown that the recession velocities of distant galaxies are only
apparently superluminal, due to the adopted definition of distance in
cosmology and the effect of special-relativistic time
dilation. Alternatively defined, {\em inertial\/} velocities are
subluminal. The effect of time dilation is also responsible for
infinite distance to the particle horizon in this model. Specifically,
the distance is infinite because the proper time of a fundamental
observer moving with the speed of light does not flow, so it never
acquires a non-zero value, necessary to perform the measurement of the
distance. (It is always `too early' to send any communication
photons.)  The particle horizon exists (i.e., the distance to it is
finite) for models with a period of initial deceleration, i.e., for
which $\Omega_m > 0$.

The empty model shares all properties of the Friedman models, that are
commonly considered as an evidence for general-relativistic expansion
of space (see Section~\ref{sec:intro}). However, in the empty model
these properties are shown to be in agreement with SR and are fully
explicable as the effects of real, relativistic motions in
space. Therefore, there is at least one Friedman model, in which
expansion of space, in detachment from expanding matter, is an
illusion. Actually, there is a whole class of such models: with the
mean matter density much smaller than the critical density, and
vanishing cosmological constant. In these models (at least since some
instant of time) expansion is approximately (but with arbitrary
accuracy) kinematic, and spacetime is approximately the static
Minkowski spacetime. The empty model is an asymptotic state of any
open model with $\Omega_\Lambda = 0$. Therefore, in any such universe,
during its evolution, expanding space should somehow, mysteriously,
disappear. The proponents of expansion of space must be able to
describe this process of disappearance. The simplest scenario for
disappearing expanding space, that comes to the mind of the author, is
that it has never existed. There is neither absolute space, nor
expanding space. All that matters is the cosmic substratum and its
relative motions. A truly Buddhist enlightenment.

\section*{Acknowledgments}
This research has been supported in part by the Polish State Committee
for Scientific Research grant No.~1 P03D 012 26, allocated for the
period 2004--2007.

\appendix
\section{Calculation of the redshift in `private space'}
\label{app:redshift}
As mentioned in Section~\ref{sec:intro}, for all Friedman models the
cosmological redshift is an accumulation of the infinitesimal Doppler
shifts caused by relative motions of closely spaced fundamental
observers along photons' trajectory. In general, this accumulation
does not sum up to a {\em global\/} Doppler shift, i.e., due to solely
the relative motion of the source and the observer. The reason for
this is that in a non-empty universe photons also undergo a
gravitational shift (e.g., Peacock 1999). In
section~\ref{sec:redshift} we have shown that in the empty model, the
cosmological redshift {\em is\/} a global Doppler shift. To do this,
we have used the special-relativistic formula for the Doppler effect,
Equation~(\ref{eq:redshift}), and shown that it leads to the correct
expression for the redshift. (That is, the same as obtained in this
model from general Eq.~\ref{eq:red}.) In this Appendix we will {\em
derive\/} this equation, summing up the infinitesimal Doppler shifts
of photons passing between neighbouring fundamental
observers. Equation~(\ref{eq:redshift}) involves the {\em inertial\/}
relative velocity of the emitter and the observer. Therefore, in our
calculation we will use {\em global\/} Minkowskian coordinates. 

At the origin of the coordinate system, let's place a source of
radiation. We will thus perform our calculations in `private space' of
the source. (The calculation in `private space' of the observer is
similar.) At time $t_e$ the source emits photons, which at time $t$
reach a fundamental observer (FO1) moving with velocity $v$, such that
\be 
v t = c (t - t_e) .
\label{eq:dist}
\ee 
Time $t$ is measured in the global inertial frame of the source, i.e.,
it is measured by an infinite set of synchronized clocks (with that
at the origin), remaining in rest relative to it. In the rest frame
of the observer FO1, a neighbouring (the one more distant from the
source) fundamental observer (FO2) moves with infinitesimal velocity
$\Delta v'$, hence an infinitesimal (so non-relativistic) Doppler
shift is
\be 
\frac{\Delta\nu'}{\nu'} = - \frac{\Delta v'}{c},
\label{eq:nonrel}
\ee 
where $\nu'$ is the photons' frequency at FO1. The velocity of FO1
relative to the source is $V = r/t$, where $r$ is its distance from
the source at time $t$. Similarly, the velocity of FO2 relative to the
source is $v = (r + \Delta r)/t$. According to the relativistic law of
composition of velocities, the velocity of FO2 relative to FO1 is
\be
\Delta v' = \frac{v - V}{1 - v V/c^2} = 
\frac{\Delta v}{1 - v^2/c^2} + \calO[(\Delta r)^2] , 
\label{eq:Dv'}
\ee
where $\Delta v \equiv v - V = (dv/dt)\Delta t$. From
Equation~(\ref{eq:dist}), $d v/d t = c\, t_e/t^2$. Furthermore, $1 -
v^2/c^2 = (t_e/t)(2 - t_e/t)$, hence
\be
\Delta v' = \frac{c \Delta t }{2 t - t_e} , 
\label{eq:Dv'2}
\ee
or
\be
- \int_{\nu_e}^{\nu_o} \frac{{\rm d} \nu'}{\nu'} = 
\int_{t_e}^{t_o} \frac{{\rm d} t}{2 t - t_e} .
\label{eq:int}
\ee
Integration yields 
\be
\ln\frac{\nu_e}{\nu_o} = \frac{1}{2} \ln\left(2 t_o/t_e - 1\right) ,
\label{eq:ln}
\ee
or, finally,
\be
\frac{\nu_e}{\nu_o} = (2 t_o/t_e - 1)^{1/2} .
\label{eq:fin}
\ee
Equation~(\ref{eq:fin}) exactly coincides with
Equation~(\ref{eq:1+z}), which, in turn, is a direct consequence of
Equation~(\ref{eq:redshift}). 

As a corollary, we can calculate an accumulation of the infinitesimal
Doppler shifts either in the local coordinates of fundamental
observers (`public space'), or in the global Minkowskian coordinates
of any selected FO (his `private space'). The local coordinates of
fundamental observers are actually more convenient, because in them,
instead of Equation~(\ref{eq:Dv'}), we have simply $\Delta v' = H'
\Delta r'$. Whatever is our choice, however, we obtain the same
result. Moreover, a calculation in `private space' is necessary to
provide a proper physical interpretation of this result: a global
Doppler shift.

\end{document}